\documentstyle[11pt]{article}

\title{\bf Cumulative particle production as a rare event}
\author{M.A.Braun\\
Dep. High-Energy Phys., Univ. of S.Petersburg, Russia\\
J.Dias de Deus\\
Dep. de Fisica/CENTRA,IST, Lisbon, Portugal\\
and C.Pajares\\
Dep. Fisica de Particulas, Univ. of Santiago de Compostela, Spain}
\date{}
\def\beq{\begin{equation}}
\def\eeq{\end{equation}}
\def\noi{\noindent}
\begin{document}
\maketitle
\medskip
\vspace{1 cm}
\begin{center} {\bf Abstract}
\end{center}
 The generalization of the Glauber formula for
cumulative production events is derived. On its basis the multiplicity
distribution in such events is related to the one in the
minimum bias events. As compared to the rare events of type $C$, the formula
involves a shift in the arguments determined by the multiplicity from a 
collision with a cluster of several nucleons.

\vspace{3.0cm}
\noi{\Large\bf hep-ph/9907303}\\
\noi{\Large\bf US-FT/??-99}
\newpage

\section{Introduction.}
In [1,2] a simple formula was derived which relates the multiplicity
distribution 
$P_C(n)$ in a rare event of the so-called type $C$ to the minimum bias (MB)
multiplicity distribution $P(n)$:
\beq
P_C(M)=\frac{MP(M)}{\langle M\rangle}
\eeq
Remarkably the right-hand side of (1) does not depend on a particular nature of
the rare process. So (1) gives a universal probability for rare events,
independent of the nature of the trigger process. This
universal formula was tested in several different rare processes for hh and hA
collisions like W and Drell-Yan dilepton pair production.  

In this paper we would like to discuss the possibility of applying (1) to
cumulative events in hA collisions. In deriving (1) it
was assumed that the geometric properties of the hA collision in the rare event
were the same as for MB events. From this point of view production of cumulative
particles, that is, in the kinematical region outside the one allowed for the
scattering on a nucleon  at rest, is different, since it involves collisions
with two or more nucleons at once. Thus it is not evident that (1) is applicable
to cumulative events, although they are certainly rare, experimentally.

 Our results show that under certain reasonable
approximations one obtains an expression for the multiplicity distribution
in a cumulative event quite similar to (1), except for a certain shift in
the arguments. The shift is due to a possible change of the mutiplicity
in a collision with a cluster of several nucleons.
In the colour string approach, clustering of nucleons leads to fusion of
strings and their percolation, which tends to diminish the multplicity [3,4]
The mentioned shift in the arguments can be observed experimentally and
thus can serve as a  signature of string fusion, once the multiplicity
distributions in cumulative  and MB events are compared.

\section{The Gribov-Glauber formula for cumulative events}
To study the multiplicity distribution in a cumulative event, one needs 
a suitable generalization of the standard Glauber formula for hA collisions.
The derivation of the standard Glauber hA amplitude for high energies
made by V.Gribov [5], expresses the hA amplitude via hN amplitudes with the
nucleon at rest. The corresponding kinematical region then evidently coincides 
with the one for the scattering on a nucleon at rest. Cumulative events, in
contrast, involve collisions with several nucleons at once. Such multinucleon
collisions were considered earlier in relation to the loop contribution in
nuclear-nuclear collisions [6] or string fusion [7]. It is remarkable, however,
that one can naturally introduce them strictly following the Gribov approach,
 provided one takes into account a more general structure for the high-energy
part of the hA amplitude. Here we only present the main steps and the final
formula. One can find some details in the Appendix.

The basic novelty in deriving the Glauber amplitude for cumulative events is
taking into account that the high energy part $H$ of the hA scattering
amplitude (see Figure), apart from the usual terms which depend on all momenta
transferred to the nucleus $q_i$ separately, may include other terms which
depend on their partial sums, say, $q_1+q_2$, $q_3+q_4+q_5$ etc. As we shall
see, such terms describe simultaneous interaction of the projectile with 
clusters of two, three etc. nucleons. 
 Correspondingly, we shall call these sums of the
transferred momenta clusters for brevity. In the general case, clustering
implies a subdivision of all transferred momenta $q_1,....q_n$ into
various groups (clusters) of one, two or more momenta. The high-energy
part can be split into terms which depend only on the total momentum
of each cluster. Some of the clusters may contain equal number of
nucleons, say, $q_1+q_2$ and $q_3+q_4$ with two nucleons each.
The order of such clusters has no influence on the resulting
contribution to the amplitude. For this reason it is convenient
to characterize the clustering structure by  occupation
numbers $\nu_k$, $k=1,2,....$ telling the number of clusters containing
$k$ nucleons (that is, $k$ transferred momenta $q_j$). Evidently
for the $n$-fold interaction with the nucleons one should have
\beq
\sum_kk\nu_k=n
\eeq
and, of course, all $\nu_k$ with $k>n$ are zero. The standard situation
when the high-energy part depends on all $q_j$ separately corresponds
to $\nu_1=n$ and $\nu_k=0$ for $k>1$.

Taking the possibility of clustering into account, we present the high-
energy part as a sum of contribution from all  possible clusters:
\beq
H(q_1,...q_n)=\prod_k\sum_{\nu_k}S_{\nu_k}H_{\nu_k}(q_1,...q_n)
\eeq
Here $H_{\nu_k}(q_1,...q_n)=H_{\nu_k}(\kappa_1,\kappa_2,
....\kappa_{\nu})$ is supposed to depend only on the total
cluster momenta $\kappa_j$, $j=1,2,....\nu$ where
\[\nu=\sum_k\nu_k\]
is the total number of clusters. We also assume that the initial momenta
$q_j$ are distributed among the clusters in some fixed way, say, in the
order of the growing cluster size. Various redestributions are then taken
into account by the evident symmetry factor
\beq
S_{\nu_k}=\frac{n!}{\prod_k(k!)^{\nu_k}\nu_k!}
\eeq

Eqs. (3) and (4) present the essence of our generalization of the
Gribov derivation. Then, following  the standard procedure of integrating
over the longitudinal momentum transfers by means of closing the contour
around the right-hand singularities of the high-energy part
and retaining the lowest mass singularities,  one readily finds a contribution
to the hA amplitude from a given clustering $\nu_k$
\beq
i{\cal A}_{\nu_k}=S_{\nu_k}\int d^2r
\prod_k [i{a}_kT_k(r)]^{\nu_k}
\eeq
Here  $a_k$ are the scattering amplitudes for a collision of the projectile
with a cluster of $k$ nucleons at rest. Their normalization is chosen to include
an extra factor $(2s)^{-1}(2m)^{1-k}$ as compared to the standard relativistic
one, where $m$ is the nucleon mass and $s$ the c.m energy squared. With this
normalization for $k=1$ (and for the total hA amplitude) twice the imaginary
part directly gives the total cross-section. The generalized profile functions
$T_k$ are defined as
\beq
T_k(r)=\int dz\rho^k(r,z)dz
\eeq
where $\rho(r,z)$ is the 3-dimensional nuclear density. As expected, they
give  the probability to find $k$ nucleons  at the same
point in the transverse plane. 

To find the total hA amplitude we have to sum (5) over all cluster
structures for fixed $n$ and then sum over all $n$ with the standard
factor $C_A^n$. This reduces to summing over all $\nu_k$ with the only
restriction
\beq\sum_kk\nu_k\leq A\eeq
In this way we obtain the hA amplitude in the form
\beq
{\cal A}=\int d^2r{\cal A}(r)
\eeq
where the hA amplitude for fixed impact parameter is given by
\beq
i{\cal A}(r)=\sum_{\nu_k}\frac{A!}
{(A-\sum_kk\nu_k)!\prod_k\nu_k!}\prod_k [i{a}_kT_k(r)/k!]^{\nu_k}
\eeq
and the summations are restricted by condition (7).

The standard Glauber formula is obtained if all $\nu_k=0$ for $k>1$.
In the general case we cannot do the summations explicitly for arbitrary
$A$. However they are easily done for a very heavy nucleus with $A>>1$.
In this limit we can drop  condition (7) and approximately take
\[\frac{A!}{(A-\sum_kk\nu_k)!}\simeq \prod_kA^{k\nu_k}\]
Then the summation over each $\nu_k$ leads to an exponential factor
and we obtain
\beq
i{\cal A}(r)=\exp \left(\sum_ki{a}_k\frac{A^kT_k(r)}{k!}\right)-1
\eeq
The subtracted unity corresponds to a term with all $\nu_k=0$ and
thus without any interaction. Evidently, neglecting all amplitudes
with $k>1$ one obtains the standard Glauber expression for a heavy
nucleus.

Note that $T_k\sim A^{1/3-k}$ and so $A^kT_k\sim A^{1/3}$. As a result
all terms in the exponent in (10) have the same order $A^{1/3}$
independent of the number of nucleons in the cluster $k$. This means that
the relative weight of the multinucleon interactions is essentially
independent of $A$. Their smallness is determined not by the large
nuclear volume as a whole but by the large internucleon distance $R_0$
in the nucleus as compared to the strong interaction radius $R_s$, which
determines the magnitude of $a_k$.
Indeed $A^kT_k\sim A^{1/3}R_0^{1-3k}$. So the relative weight of the
interaction with $k$ nucleons at once is of the order of the
dimensionless parameter $a_kR_0^{1-3k}\sim (R_s/R_0)^{3k-1}$.
Taking  $R_s\sim 1/m_{\rho}\sim 0.27 f$
one finds that raising the number of nucleons in the cluster by unity
leads to a damping factor of the order $(1/4)^3=1/64$

To calculate the contribution of multinucleon interactions one has to
know the  multinucleon amplitudes $a_k$.  They only weakly depend on the
energy $s$.
Although being on-mass-shell
quantities, they do not seem to be directly measurable. An obvious way
is to determine them from simpler processes, like the scattering on the
deuteron, triton etc, and then use the found values
for studying the cumulative kinematical region in interactions with other
nuclei.

\section{The AGK rules, cross-sections and multiplicities}
To simplify our notations we define $t_k(r)=T_k(r)/k!$ 
 Then at fixed $r$
\beq
i{\cal A}(r)=\sum_{\nu_k}\frac{A!}
{(A-\sum_kk\nu_k)!\prod_k\nu_k!}\prod_k [ia_kt_k(r)]^{\nu_k}
\eeq
The twice imaginary part of this amplitude gives the total
cross-section on the nucleus at fixed $r$
\beq
\sigma_A^{tot}(r)=-i{\cal A}-(i{\cal A})^*=
\sum_{\nu_k}\frac{A!}
{(A-\sum_kk\nu_k)!\prod_k\nu_k!}\prod_k t_k(r)^{\nu_k}
[-(ia_k)^{\nu_k} -{(ia_k)^*}^{\nu_k}]
\eeq

To obtain the AGK rules we have to split this into terms corresponding
to a given number of inelastic collision with the nuclear clusters.
We use a trivial identity
\[ ia_k+(ia_k)^*+2{\rm Im}\,a_k=0\]
It demonstrates that for a given
discontinuity of the diagram (cutting of the diagram)
each interaction may be either
cut or taken uncut on both sides of the cut.
We present in (12)
\beq
-(ia_k)^{\nu_k} -{(ia_k)^*}^{\nu_k}=
(ia_k+(ia_k)^*+2{\rm Im}\,a_k)^{\nu_k}
-(ia_k)^{\nu_k} -{(ia_k)^*}^{\nu_k}
\eeq
Terms proportional to $(2{\rm Im}\,a)_k)^{\xi_k}$ on the right-hand side
correspond to $\xi_k$-fold inelastic collision with clusters $k$.
They are
\[C_{\nu_k}^{\xi_k}(2{\rm Im}\,a_k)^{\xi_k}(ia_k+(ia_k)^*)^{\nu_k-\xi_k}
=C_{\nu_k}^{\xi_k}\sigma_k^{\nu_k}(-1)^{\nu_k-\xi_k}\]
where we denoted
\[\sigma_k=2{\rm Im}\,a_k\]
Terms without $2{\rm Im}\,a_k$ represent the diffractive part of the
cross-section
\[(ia_k+(ia_k)^*)^{\nu_k}-(ia_k)^{\nu_k} -{(ia_k)^*}^{\nu_k}\]

Summing over all sets of clusters we get the cross-section
at a given $r$ corresponding to $\xi_k$ inelastic interactions with
cluster $k$ in the form
\beq
\sigma_A^{(\xi_k)}(r)=
\sum_{\nu_k}\frac{A!}
{(A-\sum_kk\nu_k)!\prod_k\nu_k!}\prod_k C_{\nu_k}^{\xi_k}
(\sigma_kt_k(r))^{\nu_k}
(-1)^{\nu_k-\xi_k}
\eeq
where of course $\nu_k\geq\xi_k$. Changing the summation variable
$\nu_k\rightarrow\nu_k-\xi_k$ we rewrite this as
\beq
\sigma_A^{(\xi_k)}(r)=
\frac{A!}
{(A-\sum_kk\xi_k)!\prod_k\xi_k!}\prod_k(\sigma_kt_k(r))^{\xi_k} 
\sum_{\nu_k}
\frac{(A-\sum_kk\xi_k)!}{(A-\sum_kk\xi_k-\sum_kk\nu_k)!\prod_k\nu_k!}
(-\sigma_kt_k(r))^{\nu_k}
\eeq
This formula expresses the AGK rules  in the general case.
It can easily be simplified for a heavy nucleus when in the same
way as in deriving (10) we easily obtain
\beq
\sigma_A^{(\xi_k)}(r)=
\prod_k\frac{(\sigma_kA^kt_k(r))^{\xi_k}}{\xi_k!} 
\exp \left(-\sum_k\sigma_kA^kt_k(r)\right)
\eeq
in an obvious generalization of the well-known formula for collisions
with separate nucleons inside the nucleus.

The multiplicity $\mu_k$ observed in an inelastic collision with a cluster
of $k$ nucleons is not known {\it apriori}. Its magnitude depends on the
dynamics of the interaction with the cluster. In a model in which this
interaction is realized via creation and decay of colour strings,
the multiplcity $\mu_k$ depends on the interaction between the strings.
With no interaction, the nucleons, although located at the same longitudinal
point, produce strings independently so that $\mu_k=k\mu_1$.
In the opposite case all strings fuse into a pair and $\mu_k=\mu_1$.

To find the multiplicity in the event with $\xi_k$ inelastic
interactions with  cluster $k$ one has first to find the corresponding
inclusive cross-section. This corresponds to substituting in (15) one of
the cross-sections $\sigma_k$ by $\mu_k\sigma_k$ and then summing over all
such substitutions. The multiplicity is obtained by dividing the result
by the cross-section (15) itself. This gives a simple formula
\beq
\mu^{(\xi_k)}=\sum_k\xi_k\mu_k
\eeq

\section{Rare events.}
As discussed, the probability for the nucleons to form a cluster is rather
small. So in the cumulative event, in which at least one cluster with a
given $k>1$ interacts with the projectile, in
the first approximation one can neglect all configurations with  two
or more clusters with $k>1$. We shall assume that the nucleus has a
sharp edge and a constant density inside, so that
\beq
\rho(r,z)=(1/V_A)\theta(R_A^2-r^2-z^2)
\eeq
where $R_A=A^{1/3}R_0$ is the nuclear radius and $V_A=AV_0$ is the
nuclear volume. Then we find 
\beq
t_k(r)=\frac{T(r)}{k!V_A^{k-1}}
\eeq
with $T(r)\equiv T_1(r)$ the standard nuclar profile function.

With these simplifications and neglecting the contribution of
clusters in the sum over $\nu_k$ in (15) we find the cross-section
for a single inelastic interaction with a cluster $k$ and $m-1$
interactions with separate nucleons at a given impact parameter $r$
as
\beq
\sigma_A^{(m,k)}=\frac{A!}{(A-m-k+1)!(m-1)!}
\frac{\sigma_k}{k!V_A^{k-1}}(\sigma T(r))^{m}(1-\sigma T(r))^{A-m-k+1}
\eeq
For a heavy nucleus this transforms into
\beq
\sigma_A^{(m,k)}=\frac{1}{(m-1)!}
\frac{\sigma_k}{k!V_0^{k-1}}(\sigma AT(r))^{m}e^{-\sigma AT(r)}
\eeq

Let us compare this cross-section with the one for $m$ inelastic
interactions with isolated nucleons (without clustering)
\beq
\sigma_A^{(m)}=\frac{A!}{(A-m)!m!}
(\sigma T(r))^{m}(1-\sigma T(r))^{A-m}
\eeq
which for a heavy nucleus goes into
\beq
\sigma_A^{(m)}=\frac{1}{m!}
(\sigma AT(r))^{m}e^{-\sigma AT(r)}
\eeq
The comparison of (21) and (23) gives a relation
\beq
\sigma_A^{(m,k)}=\alpha_k m\sigma_A^{m}
\eeq
where we have defined
\beq
\alpha_k=
\frac{\sigma_k}{k!V_0^{k-1}}
\eeq
Eq.(24) means that the probability $P_{k}(m)$ to find apart from a
single interacting cluster
$k$ also $m-1$ interacting separate nucleons, $m=1,2,...$,  is
related to the probability $P(m)$ to find  $m$
interacting separate nucleons and no clusters  at all according to
\beq
P_{k}(m)=c_k\alpha_k mP(m)
\eeq
Since the total probability to have any number of separate interactions with
the nucleons apart from an interaction with the cluster is unity, the
sum of (26) over all $m$ should give unity, which determines the constant
$c_k$. We then obtain
\beq
P_{k}(m)=\frac{mP(m)}{\sum_nn P(n)}=
\frac{mP(m)}{\langle m\rangle}
\eeq
This relation has  the same form (1) as obtained in [1,2] for
rare events of the so-called $C$-type.
It is remarkable that the relation is independent of the
cluster characteristics ($k$ and $\alpha_k$),
the only requirement being that the interaction with it should be rare.

However the resulting observable consequences will be different.
The point is that passing to observable multiplicities, we have to
introduce the multiplicity for the interacting cluster $\mu_k$. Then the
total multiplicity $M$, observed in the event in which, apart from the
cluster, $m-1$ isolated nucleons interact will be
\beq M=\mu_k+(m-1)\mu=m\mu+\mu_k-\mu\equiv m\mu+\Delta_k\eeq
where $\mu\equiv\mu_1$ is the multiplicity for the interaction with a
isolated nucleon. So, in terms of the multiplicities, the relation (27)
implies
\beq
P_k(M)=\frac{(M-\Delta_k)P(M-\Delta_k)}{\langle M-\Delta_k\rangle}
\eeq
This is our final result. It shows that, applied to rare cluster interactions,
the approach of [1,2] leads to a very similar relation, in which however
the argument is shifted on the right-hand side, the shift depending on
the multiplicity for the given cluster. As mentioned, the latter depends
on the assumed dynamics of colour strings in the cluster. Without
any interaction between them $\mu_k-\mu=(k-1)\mu$ and the shift is
maximal. In the opposite case when  the cluster behaves in the same
manner as a single nucleon $\mu_k=\mu$ and there is no shift at all.
Thus experimental studies of the relation (29) can shed light on the
strength of the interaction between colour strings and the probability
of their fusion and percolation.

In terms of  averages $\langle\rangle_k$ for a cumulative event
involving a cluster with $k$ nucleons and
$\langle\rangle$ for MB events, one gets from (29) for the average multiplicity
\beq
\langle M\rangle_k=\frac{\langle M^2\rangle}{\langle M\rangle}+\Delta_k
\eeq
and for its dispersion squared
\beq
D^2_k=\langle M^2\rangle_k-\langle M\rangle_k^2=
\frac{\langle M^3\rangle}{\langle M\rangle}-
\frac{\langle M^2\rangle^2}{\langle M\rangle^2}
\eeq
One observes that the dispersion of the multiplicity does not depend on the
the cluster properties and is the same as for rare events of the
type $C$ found in [1,2]. As discussed there it is smaller than for MB events,
so that production of cumulative particles triggers sharpening of the
multiplicity distribution. On the other hand, the average multiplicity
in the cumulative event generally results still greater than for  rare events
of the type $C$ (given by the first term in (30)) due to the extra term 
$\Delta_k$. Only in the limitiung case when $\Delta_k=0$, which in the string
language corresponds to a very strong fusion, the two multiplicities
coincide, remaining greater than for MB events.

\section{Conclusions}
The standard Glauber-Gribov derivation of the hA-amplitude in terms of hN
amplitudes has been naturally extended to include cumulative effects.
In this way the hA amplitude is obtained in terms of the amplitudes for the
scattering of the
projectile hadron off  clusters of $k$ nucleons, $k=1,2,...$. The resulting
formula has the same structure as the original Glauber formula. By means of the
AGK cutting rules cross-sections for a given number of inelastic collisions
with clusters of $k$ nucleons are found. Assuming a collision with a cluster of
more than one nucleon to be a rare event and thus neglecting contributions
from two and more collisons with such clusters, the assocated
multiplicity distribution  is expressed in terms of the MB multiplicity
distribution. The obtained formula has the same universal form  as (1),
except for a shift $\Delta_k$ in the arguments. The shift is determined by the
difference between the multiplicities in the scattering with a cluster of $k$
nucleons and with a single nucleon. Observation of this shift potentially gives
a  possibility to study the multiplicity coming from a cluster of nucleons.
Although this does not seem to be easy experimentally, the shift could be
studied in the forthcoming experiments at RHIC and LHC, using, for instance,
the BRAHMS detector.

\section{Acknowledgemnts}
This study was supported by the NATO grant CRG. 971461. C.P.is also thankful
to CICYT (Spain) for the financial support under the contract AEN 96-1673.

\section{Appendix. The hN amlitude with multinucleon interactions}

The hA amplitude with several ($n$) rescatterings can be represented
by a generic diagram shown in Figure.  It divides into a part
related to the structure of the nucleus and the high-energy part
represented by the blob $H$. Separation of the nuclear part is, in fact,
standard and follows the original approach of V.Gribov [5]. We briefly repeat it
for self-consistency and to introduce the necessary notations.
The latter are as follows. We denote the nucleus momentum as $Ap$ and
take the nucleus at rest: ${\bf p}=0$.
The nucleon momenta before (after) the interaction
are denoted $k_i$ ($k'_i=k_i+q_i$).
The spectators correspond to $i=n+1,...A$. For them $k_i=k'_i$ and $q_i=0$.
Also $\sum q_i=0$. The projectile momentum is denoted $l$.

The expression for the amplitude corresponding to the diagram in the Figure is
\beq
i{\cal A}=\int\prod_{j=2}^{n}\frac{d^4k_j}{(2\pi)^4}
\frac{d^4k'_j}{(2\pi)^4}P(k_j)P(k'_j)\prod_{j=n+1}^{A}\frac{d^4k_j}
{(2\pi)^4}P(k_j)
i\Gamma(k_i)i\Gamma(k'_i)iH(l,k_j,k'_j)
\eeq
In this expression $H$ is the above mentioned high-energy part and
$P(k)$ is the propagator of the nucleon with momentum $k$:
\beq
P(k)=\frac{-i}{m^2-k^2-i0}\simeq\frac{-i}{\alpha^2+{\bf k}^2-2mk_0-i0}
\eeq
where $m$ is the nucleon mass and $A\alpha^2/(2m)$ is the nucleus binding
energy.
The vertex $\Gamma$ describes the transition
of the nucleus into $A$ nucleons.
We have chosen the momenta of the first nucleon $k_1$ and $k'_1$
as dependent variables, although in future we shall use a different choice
also. Evidently $k_1=-\sum_{i=2}^Ak_i$ and similarly 
$k'_1=-\sum_{i=2}^{A}k'_i$.

Standardly one starts by the integration over the zero components of the
momenta.
Since the poles coming from the  propagators of the first nucleon all lie
in the upper half-plane, we can integrate over $k_{i0}$ or $k'_{i0}$,
$i=2,...A$, just taking the residue at the pole of the corresponding
propagator
$P(k_i)$ or $P(k'_i)$. Each such integration provides a factor $2\pi/(2m)$.
The two propagators of the active nucleon in the initial and final state
together with the factors $i\Gamma(k_i)i\Gamma(k'_i)$ combine
into a product of two nuclear wave functions
\[(2m(2\pi)^3)^{A-1}\phi(k_i)\phi(k'_i)\]
so that the expression for the amplitude (32) becomes
\beq
i{\cal A}=(2m(2\pi)^3)^{A-1}\int\prod_{j=2}^{n}\frac{d^3k_j}{2m(2\pi)^3}
\frac{d^3k'_j}{2m(2\pi)^3}\prod_{j=n+1}^{A}\frac{d^3k_j}{2m(2\pi)^3}  
\phi(k_i)\phi(k'_i)iH(l,k_j,k'_j)
\eeq

The next step is to pass to the coordinate space to simplify the
dependence on the wave functions.
We present
\beq
\phi(k_i)=\int\prod \frac{d^3r_i}{(2\pi)^{3/2}}\psi(r_i)\exp(-i\sum k_jr_j)
\eeq
and then integrate over  the transverse momenta.
 The high-energy part $H$  does not
depend on the transverse momenta of the nucleons in the high-energy limit.
So the integration over them  is trivial, since all
the relevant dependence is concentrated in the exponentials.
We readily obtain
\[
i{\cal A}=\int\prod_{j=1}^{A-1}\frac{dk_{zj}}{2\pi}
\prod_{j=2}^{n}\frac{dq_{zj}}{2m(2\pi)}
\]\[iH(l_z,k_{jz},q_{jz})  
\prod dz_i\exp(-i\sum k_{zj}(z_j-z'_j))
\prod dz'_i\exp(i\sum q_{zj}(z'_j-z'_1))\]\beq
d^2r_1\prod_{j=n+1}^{A-1}d^2r_j
\psi(r_1=r_2=r_3+...r_{n},r_j; z_i) 
\psi(r_1=r_2=r_3+...r_{n},r_j; z'_i) 
\eeq
As expected on the physical grounds,
the $n$ nucleons which take part in the
rescattering  have to be taken at the same impact parameter.

Now we have to integrate over the longitudinal momenta.
Evidently the
integrand does not depend on the $k_z$ of the spectators. So these
interactions are done trivially and add  a factor
\[ \prod_{i=n+1}^{A-1}2\pi\delta(z_i-z'_i)\]
Apart from the factor $(2\pi)^{A-n-1}$, these $\delta$ functions 
 convert the double integration over $z$  into a single one
for the spectators.
Together with the integration over their transverse coordinates this
turns the product of the wave functions into the nuclear $\rho$-matrix
for $n+1$ nucleons taking part in the interaction.

The high-energy part $H$ does not depend on $k_{jz}$ but only on
$q_{jz}$. This property follows from the relativistic invariance, due to
which $H$ depends on products like $lk_j$ and $lq_j$. With small spatial
componets of $k_j$, the products of the first tipe all are approximately
equal to $lp$ and only the second are real variables.  So one can also
trivially integrate over $k_{jz}$, $j=1,...n$, which gives a factor
$(2\pi)^n$ and puts $z_j=z'_j$, $j=1,...n$
We get
\beq
i{\cal A}=\int\prod_{j=2}^{n}\frac{dq_{zj}}{2m(2\pi)}
iH(l_z,q_{jz}) \prod dz_i\exp(i\sum q_{zj}(z_j-z_1))
d^2r\rho(rz_1,rz_2,....rz_n|rz_1,rz_2,...rz_n)
\eeq
At this point all nuclear effects are taken into account by
 the nuclear $\rho$-matrix taken at interaction points
$(r,z_j)$.  The final task  is to integrate over the transferred
longitudinal momenta $q_{jz}$. To do this we take into account the generalized
structure of the high-energy part as a function of the transferred momenta,
discussed
in Sec. 3 and presented in Eqs (3) and (4).

Let us find the contribution to the hA amplitude with $n$ interactions (7)
from a term in (3) with given $\nu_k$. All integrations over $q_j$ can
be divided into $\nu-1$ integrations over the total cluster momenta
$\kappa_2,...\kappa_{\nu}$ ($\kappa_1=-\sum_{j=2}^{\nu}\kappa_j$)
and the rest $q_j$, without, say, the first $q$ from each cluster.
The integration over the latters is trivial, since, the high-energy part
does not depend on them. It will make all longitudinal points $z_j$
equal within each cluster, in accordance with the intuitive definition
of a cluster or simultaneous interaction with several nucleons.

To write down the obtained expression in an understandable way,
at this point we make the usual simplifying assumption about the
structure of the nuclear $\rho$ matrix: that of absence of correlations
and consequently of factorization
\beq
\rho(rz_1,rz_2,....rz_n|rz_1,rz_2,...rz_n)=\prod_j\rho(rz_j)
\eeq
where $\rho(rz)$ is the usual nuclear density (normalized to unity).
Then the contribution to the amplitude from $H_{\nu_k}$ will be given by
\beq
i{\cal A}=S_{\nu_k}(2m)^{-n+1}\int
\prod_{j=2}^{\nu}\frac{d\kappa_j}{2\pi}
iH_{\nu_k}(\kappa_j)  
\prod_{i=1}^{\nu} dz_i\exp(i\sum \kappa_j(z_j-z_1))
d^2r\prod_{j=1}^{\nu}\rho^{k_j}(rz_j)
\eeq
where $k_j$ is the number of nucleons in the cluster at point $z_j$

This expression can now be treated in the same way as in the original
Gribov approach. Instead of $\kappa_i$ we introduce $\nu-1$ cumulative
longitudinal momentum
transfers $t_i=l\sum_{j=1}^{i}\kappa_j$. This gives a factor
$(2l)^{1-\nu}$. The Feynman
integration contour over each of $t_i$ can be
closed around the right-hand side singularities. In the spirit of the
Glauber approximation, from these singularities we retain one the pole
singularity at $t_i=0$ corresponding to the single nucleon intermediate
state. Taking the residue at this point makes the exponential factor in
(12) equal to unity. The residue of $iH_{\nu_k}$ itself is a product of
$\nu$ on-mass-shall forward connected amplitudes $ia_k$ for the
interaction of the
projectile with $k$ nucleons at rest, where $k$ is the number of nucleons in
a cluster. After that the final integration over the cluster points $z_j$
becomes trivial and leads to a standard profile function and its
evident generalization. We obtain 
\beq
i{\cal A}_{\nu_k}=S_{\nu_k}(4ml)^{1-\nu}(2m)^{\nu-n}\int d^2r
\prod_k [ia_kT_k(r)]^{\nu_k}
\eeq
where
\beq
T_k(r)=\int dz\rho^k(r,z)dz
\eeq
is an obvious generalization of the profile function.
Note that $4ml=s$ where $s$ is the standard c.m. energy squared.
It is convenient to pass to reduced amplitudes
\beq a_k\rightarrow \frac{a_k}{2s(2m)^{k-1}},
\ \ {\rm dim}\,{a_k}=1-3k\eeq
and similarly for ${\cal A}$. After that one obtains  Eq. (5).

\section{References}

\hspace*{0.6 cm}1. J.Dias de Deus, C.Pajares and C.Salgado, Phys. Lett. {\bf B
408}(1997) 417;{\bf B409} 474.

2. J.Dias de Deus and C.Pajares, Phys.Lett. {\bf B 442} (1998)  395.

3. M.A.Braun and C.Pajares, Nucl. Phys. {\bf B 390} (1993), 542; 559.

4. N.Armesto, M.A.Braun, E.G.Ferreiro and C.Pajares, Phys. Rev. Lett {\bf 77}
(1996) 3736.

5. V.N.Gribov, JETP {\bf 56} (1969) 892; {\bf 57} (1969) 1306.

6. M.A.Braun, Yad. Fiz. {\bf 45} (1987) 1625.

7. M.A.Braun and C.Pajares, Phys. Rev. {\bf C 51} (1995) 879.

\section{Figure caption}

The hA scattering amplitude. Double lines show the nucleus target.

\newpage

\section{Figure}

\setlength{\unitlength}{1.5 pt}
\begin{picture}(100,210)(0,0)
\thicklines
\put(110,75){\oval(120,30)}
\put(20,75){\line(1,0){30}}
\put(170,75){\line(1,0){30}}
\put(20,19.){\line(1,0){30}}
\put(20,21){\line(1,0){30}}
\put(170,19.){\line(1,0){30}}
\put(170,21.){\line(1,0){30}}
\put(50,20){\line(2,5){16}}
\put(50,20){\line(3,2){60}}
\put(50,20){\line(5,2){100}}
\put(170,20){\line(-2,5){16}}
\put(170,20){\line(-3,2){60}}
\put(170,20){\line(-5,2){100}}
\put(50,20){\line(1,0){120}}
\put(105,70){\large H}
\put(47,40){$k_1$}
\put(167,40){$k'_n$}
\put(82,25){$k_1$}
\put(133,25){$k_n$}
\put(20,25){$Ap$}
\end{picture}

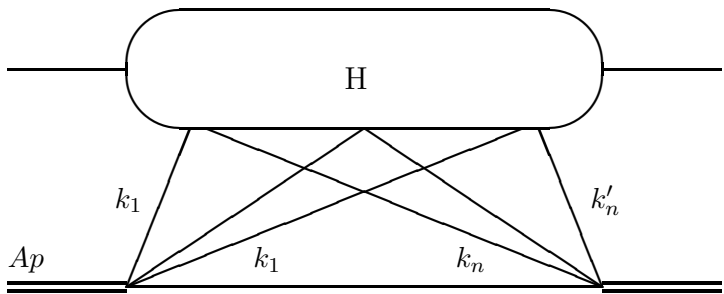
\end{document}